\documentclass[doublespacing]{elsart}

\usepackage{amssymb}

\def\max{{\mathop{\rm max}}}
\def\min{{\mathop{\rm min}}}
\def\norm#1{\left|\mkern-2mu\left|#1\right|\mkern-2mu\right|}

\begin{document}

\begin{frontmatter}

\title{Constraints on the spectral distribution of energy and
enstrophy dissipation in forced two-dimensional turbulence}

\author{Chuong V. Tran and Theodore G. Shepherd\corauthref{cor}}
\corauth[cor]{Email: tgs@atmosp.physics.utoronto.ca, Fax: (416) 978-8905}

\address{Department of Physics, University of Toronto,
60 St. George Street, Toronto, ON, Canada, M5S 1A7}


\begin{abstract}

We study two-dimensional turbulence in a doubly periodic domain driven
by a monoscale-like forcing and damped by various dissipation 
mechanisms of the form $\nu_{\mu}(-\Delta)^{\mu}$. By ``monoscale-like'' 
we mean that the forcing is applied over a finite range of wavenumbers 
$k_\min \leq k \leq k_\max$, and that the ratio of enstrophy injection 
$\eta \geq 0$ to energy injection $\varepsilon \geq 0$ is bounded by 
$k_\min^2 \varepsilon \leq \eta \leq k_\max^2 \varepsilon$. Such a forcing is 
frequently considered in theoretical and numerical studies of 
two-dimensional turbulence. It is shown that for $\mu\geq 0$ the asymptotic 
behaviour satisfies
\begin{eqnarray}
\norm u_1^2&\leq&k_\max^2\norm u^2,\nonumber
\end{eqnarray}
where $\norm u^2$ and $\norm u_1^2$ are the energy and enstrophy,
respectively. If the condition of monoscale-like forcing holds only  in a 
time-mean sense, then the inequality holds in the time mean. It is also shown 
that for Navier-Stokes turbulence ($\mu = 1$), the time-mean enstrophy 
dissipation rate is bounded from above by $2\nu_1 k_\max^2$. These results 
place strong constraints on the spectral 
distribution of energy and enstrophy and of their dissipation, and 
thereby on the existence of energy and enstrophy cascades, in such 
systems. In particular, the classical dual cascade picture is shown 
to be invalid for forced two-dimensional Navier--Stokes 
turbulence ($\mu=1$) when it is forced in this manner. 
Inclusion of Ekman drag ($\mu=0$) along with molecular viscosity 
permits a dual cascade, but is incompatible with the log-modified 
$-3$ power law for the energy spectrum in the enstrophy-cascading 
inertial range. In order to achieve the latter, it is necessary to 
invoke an inverse viscosity ($\mu<0$). These constraints on permissible 
power laws apply for {\it any} spectrally localized forcing, not just for 
monoscale-like forcing.

\end{abstract}


\end{frontmatter}

\newpage
\section{Introduction}
It has long been recognized (see for example Fj\o rtoft \cite{Fjortoft}) 
that the simultaneous existence of two quadratic inviscid invariants, 
energy and enstrophy, drastically changes the picture of
two-dimensional (2D) Navier--Stokes turbulence in comparison with that 
of its 3D counterpart. In the 1960's it was suggested by Kraichnan 
\cite{Kraichnan1}, Leith \cite{Leith}, and Batchelor \cite{Batchelor} 
(hereafter referred to as KLB) that these two constraints give
rise to the realization of two distinct inertial ranges in wavenumber 
space for forced 2D turbulence. More precisely, the KLB theory
predicts that for a 2D Navier--Stokes fluid driven by a spectrally 
localized forcing, energy is transferred to smaller wavenumbers while
enstrophy is transferred to larger wavenumbers. For a system of finite 
size the energy is then predicted to cascade to the long-wavelength 
end of the spectrum up to the largest scale available (the system's 
linear length scale), while the enstrophy is predicted to cascade to 
the other end of the spectrum, down to a dissipation length scale. 
This prediction is well confirmed numerically in transient evolution 
from spectrally localized initial conditions. An immediate corollary 
of the dual cascade and the hypothesis of a scaling symmetry is that 
the energy spectrum scales as $k^{-5/3}$ in the energy-cascading range 
and as $k^{-3}$ (with a possible logarithmic correction) in the 
enstrophy-cascading range, where $k$ is the wavenumber. The latter 
scaling is consistent with the hypothesis that the dissipation of
enstrophy is confined to the small scales; however, the former scaling 
is inconsistent with the hypothesis that the dissipation of energy is 
confined to the large scales. Hence, these scaling laws are incompatible 
with a persistent dual cascade in finite systems. This suggests that 
there is a problem with the KLB theory applied to finite systems. 
(We refer here only to the 
forced-dissipative equilibrium behaviour, not to transient evolution 
from spectrally localized initial conditions.)

For the case of a time-independent force, Constantin, Foias and Temam
\cite{Constantin3,Constantin4} have proven the existence of a compact global 
attractor for forced 2D Navier--Stokes turbulence in a bounded
domain. Constantin, Foias and Manley \cite{Constantin2} have
furthermore shown that for the special case of a doubly periodic
domain and forcing of a single length scale, the KLB scaling laws
cannot be achieved on the global attractor. Although this strong
result appears to prove the unrealizability of the KLB theory for
forced 2D Navier--Stokes turbulence in a finite domain, as argued heuristically 
in the 
previous paragraph, the question arises whether it is an artifact of 
the special choice of a constant monoscale forcing.

In this paper we extend the results of Constantin, Foias and Manley 
\cite{Constantin2} in several directions. First, for their choice of 
a constant monoscale forcing we derive an asymptotic bound on the enstrophy in 
terms of the energy and show that the time-mean enstrophy dissipation occurs 
around the forcing 
scale, precluding the existence of an enstrophy-cascading inertial 
range (whatever the energy spectrum). This result applies for any 
dissipation operator of the form $\nu_{\mu}(-\Delta)^{\mu}$ with 
$\mu\geq 0$, not just molecular viscosity. Second, we show that the 
result applies for any ``monoscale-like" forcing, by which we mean 
forcing over a range of wavenumbers $k_\min \leq k \leq k_\max$ where 
the ratio of enstrophy injection $\eta \geq 0$ to energy injection 
$\varepsilon \geq 0$ is bounded by 
$k_\min^2 \varepsilon \leq \eta \leq k_\max^2 \varepsilon$ 
for all possible velocity fields $u$. 
This is a classical (though not exclusive) scenario for the KLB theory 
(e.g. Kraichnan \cite{Kraichnan1}, p.1421b; Pouquet et al. \cite{Pouquet}, 
p.314; and Lesieur \cite{Lesieur}, p.291), and is furthermore a common set-up in  
numerical simulations of forced 2D turbulence (e.g. Lilly \cite{Lilly}; 
Basdevant et al. \cite{Basdevant}; Shepherd \cite{Shepherd}).
\footnote{Constantin, Foias and Manley \cite{Constantin2} showed that 
the KLB scaling is potentially realizable (not to say it actually
occurs) if the forcing simultaneously injects energy at higher wavenumbers 
and removes energy at lower (though possibly nearby) wavenumbers. In 
this case the ratio of
enstrophy to energy injection can be much larger than the square of 
the characteristic forcing wavenumbers and could actually lie in the 
dissipation range. Such a forcing is not monoscale-like by our definition.} If 
the condition of monoscale-like forcing holds only in a time-mean sense, then 
the time-mean results still go through and the asymptotic bound is replaced 
by a bound involving the time-mean energy and enstrophy.

The question then arises whether the KLB theory and the dual cascade 
picture can be recovered, for monoscale-like forcing, with other dissipation 
mechanisms. It is
shown that the introduction of Ekman drag ($\mu=0$)---a commonly used 
numerical device and one with some physical justification for
geophysical applications \cite{Pedlosky}---in addition to regular 
viscosity permits the existence of an energy dissipation range at
large scales and an enstrophy dissipation range at small scales, thus 
allowing a dual cascade. This is consistent with numerical results, 
e.g. Boffetta et al. \cite{Boffetta}, which demonstrate an inverse 
energy cascade and a $-5/3$ power law under these conditions. However, 
the log-modified $-3$ power law of Kraichnan \cite{Kraichnan2} for 
the energy spectrum in the enstrophy-cascading range is shown to be 
unachievable; the spectrum must be algebraically steeper than $-3$. 
This result is consistent with numerical simulations of forced 2D 
turbulence under these conditions, which find steeper spectra. (e.g. 
Maltrud and Vallis \cite{Maltrud} find the spectral slope of the
enstrophy-cascading range to be between $-3$ and $-4$: the majority 
of their simulations yields values between $-3.3$ and $-3.6$. They 
also recover the scaling $k^{-5/3}$ in the energy-cascading range.) 

It is finally shown that allowing an inverse viscosity ($\mu<0$)
together with regular viscosity does permit KLB scaling. It is notable 
that the numerical simulations of Borue \cite{Borue}, and the more 
recent high-resolution numerical simulations of Lindborg and Alvelius 
\cite{Lindborg}, which both claim to exhibit a $-3$ power-law 
enstrophy-cascading range, employ an inverse viscosity. The rather
surprising implication of our results is that verifying the KLB theory 
in the case of monoscale-like forcing does not only depend on achieving 
sufficiently high Reynolds numbers 
(as is commonly believed), but also depends on the nature of the 
rather ad hoc dissipation operator employed at the large scales.

The alternative, considered by Constantin et al.
\cite{Constantin2}, is to abandon the notion of monoscale-like forcing. 
This interesting possibility is not considered here.

The remainder of this paper is organized as follows. Section 2
describes the 2D Navier--Stokes equations and their mathematical 
setting, together with some basic inequalities. The concept of monoscale-like 
forcing is discussed further, and some examples given. Section 3 reviews 
some classical ideas and arguments in 2D turbulence. After those 
preliminaries, section 4 reports an asymptotic analysis, the results 
of which include a dynamical constraint and an upper bound on the energy   
dissipation rate. Section 5 extends the results of section 4 in an 
attempt to bound the time-mean enstrophy dissipation rate and explore possible 
scaling laws for the energy spectrum. 
These results in fact apply for {\it any} spectrally localized forcing, not just 
for monoscale-like forcing. The paper 
ends with some concluding remarks in the final section.   

\section{Governing equations and basic inequalities}

We consider 2D incompressible fluid motion confined within a doubly 
periodic rectangular domain $\Omega$. The fluid is assumed to be
driven by a monoscale-like forcing (to be discussed further below) and 
damped by a variety of possible dissipation mechanisms including Ekman drag,
hypoviscosity, molecular viscosity, hyperviscosity, and inverse 
viscosity. The 2D Navier--Stokes equations which govern the fluid
motion are written in abstract form in a function space $H$ as
\begin{eqnarray}
\frac{{\rm d}u}{{\rm d}t}+B(u,u)+\sum_{\mu}\nu_{\mu}A^{\mu}u&=&f,
\label{nseqn}\\ u(t=0)&=&u_0,\nonumber
\end{eqnarray}
where $\nu_{\mu}>0$ is a generalized viscosity coefficient, 
$A\equiv-\Delta$, and $f$ is the forcing. The number $\mu$ will be called the 
degree of 
viscosity. When $\mu=1$ we have the usual molecular viscosity, while
$\mu=0$ corresponds to Ekman drag. The cases in which $\mu>1$ and
$0<\mu<1$ correspond to hyperviscosity and hypoviscosity,
respectively. We also consider inverse viscosity, i.e., negative
values of $\mu$. The summation is taken over all dissipation channels
involved. A detailed description of the functional analysis setting
for (\ref{nseqn}) is given in Constantin and Foias \cite{Constantin1} 
or Temam \cite{Temam1,Temam2}. We recall that $H$ is the $L^2$-space of
periodic, non-divergent functions with vanishing average in $\Omega$. 
$B(u,u)=P((u\cdot\nabla)u)$ where $P$ is the orthogonal projection in 
$L^2$ onto $H$. We denote by $H^{\alpha}$ the domain of definition of 
$A^{\alpha/2}$ for real $\alpha$. The (degenerate) positive eigenvalues 
of $A$ are denoted by $\lambda_k$ with the index $k$ being the
wavenumber, and the eigenspace corresponding to $\lambda_k$ is denoted
by $H(\lambda_k)$. We will occasionally refer to $\lambda_k^{-1/2}$ 
(or simply $k^{-1}$) as the length scale associated with the wavenumber~$k$.

The scalar product and the norm in $H^{\alpha}$ are given respectively by
\begin{eqnarray}
(u,v)_{\alpha}&=&\int_{\Omega}u\cdot A^{\alpha}v~{\rm d}x,\\
\norm u_{\alpha}&=&(u,u)_{\alpha}^{1/2}.
\end{eqnarray}
The cases where $\alpha=0,1$ are special and the corresponding
$H$-norm (the superscript and subscript `0' are omitted in this case) 
and $H^1$-norm are known in the literature as the energy and enstrophy 
norm, respectively. A geometric constraint, referred to as the
Poincar\'e inequality, is
\begin{eqnarray}
\norm u_{\alpha+\beta}^2&\geq&\lambda_1^{\alpha}\norm u_{\beta}^2
\end{eqnarray}
for non-negative $\alpha$, where $\lambda_1\equiv\lambda_{k_1}$, with
$k_1$ being the smallest wavenumber, is the first (smallest)
eigenvalue of $A$ in $H$. The inequality reverses direction
for non-positive $\alpha$.

The bilinear operator $B(\cdot,\cdot)$ satisfies  
\begin{eqnarray}
(Au,B(v,v))+(Av,B(v,u))+(Av,B(u,v))&=&0
\end{eqnarray}
for $u,v\in H^2$, and
\begin{eqnarray}
(u,B(v,w))&=&-(w,B(v,u))
\end{eqnarray}
for $u,w\in H^1$ and $v\in H$. These identities arise by virtue of the
non-divergent and periodic properties of the velocity field. In 
particular, we have
\begin{eqnarray}
(u,B(u,u))&=&0,\label{ortho1}\\
(Au,B(u,u))&=&0.\label{ortho2}
\end{eqnarray}
We shall collectively refer to the above identities as the
orthogonality properties of the nonlinear term. In the absence of
forcing and dissipation, they give rise to conservation of energy and 
enstrophy. These conservation laws form the foundation for the idea of
the inverse cascade of energy and the direct cascade of enstrophy, and
the associated concepts of inertial ranges, as mentioned earlier and 
reviewed in the next section.

A well-known mathematical fact is the existence of a bounded
finite-dimensional global attractor for the traditional 2D 
Navier--Stokes equations in a finite domain (as considered here). The
research performed on this subject constitutes a rich literature. The
book of Temam \cite{Temam2} provides a good treatment of the subject 
and a complete list of references. Some articles suitable for quick 
reference are Constantin et al. \cite{Constantin3,Constantin4} and Ziane 
\cite{Ziane}. We shall not attempt to demonstrate the existence of
such an attractor for the 2D Navier--Stokes system (driven by a 
time-independent forcing) with various degrees of viscosity 
(or combinations of them, as presently considered) but take its 
existence to be given.

In this study $f$ is assumed to be monoscale-like, as defined in the 
Introduction. This means that the ratio of enstrophy to energy injection, 
which we denote by $\lambda$, is bounded a priori within a certain range 
and in particular from above, for all $u$. 
The time-independent monoscale body force $f\in H(\lambda_s)$, for an
eigenvalue $\lambda_s$, considered by Constantin et al. 
\cite{Constantin2} is a special case with $\lambda=\lambda_s$, although 
it is a rather peculiar special case, for two reasons. First, the 
instantaneous energy and enstrophy injection are not constrained to be 
non-negative\footnote{For this reason, a monoscale body force is not, 
strictly speaking, a special case of monoscale-like forcing, even though our 
results apply to both cases.}, although they clearly must be so in the time 
mean since the dissipation of both quantities is non-negative. This 
means that the condition $\lambda = \lambda_s$ does  not necessarily 
extend, even approximately, to the case of a constant external forcing 
applied over a range of wavenumbers; such a forcing is not 
``monoscale-like'' in the sense we require (and as is often assumed in the 
KLB theory). Second, there exists an exact stationary solution (given by 
(\ref{StatSol})). Because of this, there appears to be a belief that the 
flow in this case collapses onto the stationary solution and is not 
turbulent. However, for sufficiently strong forcing the forced scales 
will be unstable to nonlinear interactions, and numerical simulations 
(J.C. Bowman, personal communication, 2001) indeed confirm that there 
is nothing pathological about the case of a constant monoscale forcing: 
it develops a full spectrum, albeit one constrained 
by the inequalities discussed here.

The general case of a monoscale-like forcing applied over a finite range of  
wavenumbers, as often imagined in the KLB theory, can be realized in several 
ways. One example (Shepherd \cite{Shepherd}) is 
$f=\sum_KcP(\lambda_k)u/\norm{P(\lambda_k)u}^2$ where $c>0$, 
$P(\lambda_k)$ is the projection onto $H(\lambda_k)$, $\norm\cdot$ is the energy 
norm, and the summation is over the
restricted set of wavenumbers $K$. For this forcing the energy and enstrophy
injection are both constant in time and $\lambda$ is the
mean of $\lambda_k$ over $K$. Another example is $f=\sum_KcP(\lambda_k)u$. 
Although the energy and enstrophy injection are now variable in time, both are 
positive definite and $\lambda$ lies within the range 
$[\lambda_{\min},\lambda_{\max}]$, where $\lambda_{\min}$ and 
$\lambda_{\max}$ are, respectively, the minimum and 
maximum $\lambda_k$ in $K$. This commonly used forcing is known as 
``instability'' forcing, for obvious reasons; in the geophysical context it is 
used in 2D turbulence to mimic forcing of the barotropic component of the flow 
by baroclinic instability (Lilly \cite{Lilly}; Basdevant et al. 
\cite{Basdevant}). In both these examples, $f$ depends on $u$. Another 
commonly used  forcing is white-noise forcing over $K$ (Lilly \cite{Lilly}). 
This is harder 
to control a priori, but in practice gives positive enstrophy and 
energy injection with $\lambda$ generally fluctuating within the range
$[\lambda_{\min},\lambda_{\max}]$. For this case, the condition of 
monoscale-like forcing would need to be verified a posteriori, and would apply 
at best only in a time-mean sense. The results derived here would then apply 
only in the time mean.

We now introduce some terminology and derive some preliminary
estimates which are employed later. In the theory of turbulence the 
characteristic wavenumber $\overline{k}$ defined by
$\overline{k}=\norm u_{1/2}^2/\norm u^2$ is often studied. However, we 
find it more natural and convenient in the context of this work to 
define the parameter
\begin{eqnarray}
\Lambda&=&\frac{\norm u_1^2}{\norm u^2}~,
\end{eqnarray}
the square root of which has the dimension of a wavenumber. Moreover,
let $\Lambda_{E\mu}$ and $\Lambda_{Z\mu}$ be defined (for $\mu\neq 0$) by
\begin{eqnarray}
\Lambda_{E\mu}&=&\left(\frac{\norm u_{\mu}^2}{\norm u^2}\right)^{1/\mu},\\
\Lambda_{Z\mu}&=&\left(\frac{\norm u_{1+\mu}^2}{\norm 
u_1^2}\right)^{1/\mu},\end{eqnarray}
so that $2\nu_{\mu}\Lambda_{E\mu}^{\mu}$ and 
$2\nu_{\mu}\Lambda_{Z\mu}^{\mu}$ are, respectively,
the (instantaneous) energy dissipation and enstrophy dissipation rates
due to $\nu_{\mu}A^{\mu}$. The values of $\Lambda_{E\mu}$ and
$\Lambda_{Z\mu}$ therefore indicate where, in wavenumber space, 
viscosity primarily operates in the dissipation of energy and
enstrophy, respectively, for the dissipation operator of degree
$\mu$. Note that $\Lambda_{E1}\equiv\Lambda$; we will use the
latter symbol exclusively in what follows.

We now derive several fundamental inequalities relating these
parameters. In the following, $\alpha$, $\beta$, $\gamma$, and 
$\lambda$ are real numbers, and $\phi$ is an appropriate function so
that all norms involved are well-defined. First, we have the following
interpolation-type inequality which can be shown by H{\"o}lder's inequality:
\begin{eqnarray}
\norm\phi_{\beta+\gamma}^{\alpha}&\leq&\norm\phi_{\alpha+\gamma}^{\beta}
\norm\phi_{\gamma}^{\alpha-\beta},\label{holder1}
\end{eqnarray}
for $\alpha\geq\beta\geq 0$. Note that the inequality reverses
direction for $\alpha\leq\beta\leq 0$. Second, let us define
\begin{eqnarray}
G(\alpha,\beta,\lambda,\phi)&=&\sum_{\lambda_k}
(\lambda^{\alpha}-\lambda_k^{\alpha})(\lambda^{\beta}
-\lambda_k^{\beta})\norm{P(\lambda_k)\phi}^2\label{epsilon}
\end{eqnarray}
for $\lambda>0$. It is obvious from (\ref{epsilon}) that
$G(\alpha,\beta,\lambda,\phi)$ is positive (negative) if and 
only if $\alpha\beta>0$ ($\alpha\beta<0$) and $\phi\not\in H(\lambda)$
when $\lambda$ is an eigenvalue of $A$. This implies that when
$\alpha\beta\neq 0$, $G(\alpha,\beta,\lambda,\phi)=0$ if and 
only if $\lambda$ is an eigenvalue of $A$ and $\phi\in H(\lambda)$. 
This important feature of $G(\alpha,\beta,\lambda,\phi)$ is
used in some arguments of section 4. By rearranging terms we obtain
\begin{eqnarray}
G(\alpha,\beta,\lambda,\phi)&=&\norm\phi^2_{\alpha+\beta}
-\lambda^{\beta}\norm\phi^2_{\alpha}-\lambda^{\alpha}
(\norm\phi^2_{\beta}-\lambda^{\beta}\norm\phi^2).
\end{eqnarray}
It is then easy to show that
\begin{eqnarray}
G\left(\alpha,\beta,\frac{\norm\phi^{2/\alpha}_{\alpha}}
{\norm\phi^{2/\alpha}},\phi\right)&=&\frac{\norm\phi^2
\norm\phi^2_{\alpha+\beta}-\norm\phi^2_{\alpha}
\norm\phi^2_{\beta}}{\norm\phi^2},
\end{eqnarray}
from which it follows that
\begin{eqnarray}
\norm\phi\,\norm\phi_{\alpha+\beta}&\geq&\norm\phi
_{\alpha}\norm\phi_{\beta}\mbox{~~~for~}\alpha\beta~\geq~0,\\
\norm\phi\,\norm\phi_{\alpha+\beta}&\leq&\norm\phi
_{\alpha}\norm\phi_{\beta}\mbox{~~~for~}\alpha\beta~\leq~0.
\end{eqnarray}

The ratio of $\Lambda_{E\mu}$ to $\Lambda_{E\mu'}$ is given by
\begin{eqnarray}
\frac{\Lambda_{E\mu}}{\Lambda_{E\mu'}}&=&\left[\frac{\norm
u_{\mu}^{1/\mu}\norm u^{1/\mu'-1/\mu}}
{\norm u_{\mu'}^{1/\mu'}}\right]^2,
\end{eqnarray}
which, by applying (\ref{holder1}) with $\alpha=\mu$, $\beta=\mu'$, 
and $\gamma=0$ implies
\begin{eqnarray}
\Lambda_{E\mu}&\geq&\Lambda_{E\mu'},\mbox{~~for~}\mu~\geq~\mu'~>~0,\\ 
\Lambda_{E\mu}&\leq&\Lambda_{E\mu'},\mbox{~~for~}\mu~\leq~\mu'~<~0.
\end{eqnarray}
By applying (\ref{holder1}) to the ratio of
$\Lambda_{Z\mu}/\Lambda_{Z\mu'}$ we also obtain the same
inequalities for $\Lambda_{Z\mu}$ and $\Lambda_{Z\mu'}$. The ratio of
$\Lambda_{Z\mu}^{\mu}$ to $\Lambda_{E\mu}^{\mu}$ is given by
\begin{eqnarray}
\frac{\Lambda_{Z\mu}^{\mu}}{\Lambda_{E\mu}^{\mu}}&=&1+\frac
{G(\mu,1,\Lambda,u)\norm u^2}{\norm u_1^2\norm u_{\mu}^2}.
\end{eqnarray}
Since $G(\mu,1,\Lambda,u)$ takes the sign of $\mu$ (see
above), the above ratio indicates the simple fact that a viscosity
dissipates more enstrophy than energy, while an inverse viscosity 
dissipates more energy than enstrophy. Note, however, that
\begin{eqnarray}
\Lambda_{Z\mu}&\geq&\Lambda_{E\mu},\mbox{~~~for all~}\mu~\neq~0.
\end{eqnarray}
We will compare $\Lambda$, $\Lambda_{E\mu}$ and $\Lambda_{Z\mu}$
with $\lambda$ in subsequent sections.

\section{The KLB theory}

We now review the arguments leading to the KLB theory. In the absence of 
forcing and dissipation,
the 2D Navier--Stokes system conserves energy and enstrophy due to 
(\ref{ortho1}) and (\ref{ortho2}).
There is a consensus that an initial distribution of energy cascades towards 
larger scales, with a
downscale cascade of energy practically forbidden. Arguments against the 
downscale cascade of energy
were advanced by Taylor \cite{Taylor} and Lee \cite{Lee}. A celebrated 
``proof" of the upscale cascade
is due to Fj\o rtoft \cite{Fjortoft}. In his argument, Fj\o rtoft 
considered the change in energy for
three different scales coupled nonlinearly. Because of conservation of 
enstrophy, energy must flow
from the intermediate scale to the smaller and larger scales, or vice 
versa. For energy initially on
the intermediate scale, it was argued that the larger scale acquires most of the 
transferred energy.
In the present setting Fj\o rtoft's argument goes as follows. Let $l<k<m$ be the 
three wavenumbers
(corresponding to three scales $m^{-1}<k^{-1}<l^{-1}$) that are involved in the 
energy transfer. Furthermore, let $\Delta E(\cdot)$ denote the change of energy 
in each scale. It
is easy to see from the two conservation laws that
\begin{eqnarray}\Delta E(l)+\Delta E(k)+\Delta E(m)&=&0,\nonumber\\
\lambda_l\Delta E(l)+\lambda_k\Delta E(k)+\lambda_m\Delta 
E(m)&=&0.\nonumber\end{eqnarray} 
Hence,
\begin{eqnarray}\Delta 
E(l)&=&-\frac{\lambda_m-\lambda_k}{\lambda_m-\lambda_l}\Delta
E(k),\nonumber\\\Delta 
E(m)&=&-\frac{\lambda_k-\lambda_l}{\lambda_m-\lambda_l}\Delta
E(k).\nonumber\end{eqnarray}
It is argued in \cite{Fjortoft} that if one takes $m=2k=4l$, for example, then
($\lambda_m=4\lambda_k=16\lambda_l$)
\begin{eqnarray}\frac{\Delta E(l)}{\Delta E(m)}&=&4.\nonumber\end{eqnarray}
Therefore, changes in the kinetic energy are distributed in the ratio 4:1 on the 
components with
the double and half scale, respectively, if no other components are involved in 
the energy
transformation.

There are a number of problems with this argument. First, the inviscid system is 
time reversible, so
an upscale cascade of energy cannot be established by the above argument alone. 
Second, the three
scales involved in the energy transformation must satisfy the triad 
requirement,\footnote{Although
the geometry considered in \cite{Fjortoft} is a sphere, interactions in the form 
of triads are to
be observed \cite{Silberman}. For an interacting triad of similar scales where 
one wishes to have
the smallest and largest scales as far away from the intermediate scale as 
possible, the scale
ratio of 1:2:3 is a better approximation than 1:2:4. For such a case, the ratio 
of the upscale
cascade energy to the downscale cascade energy is approximately 5:3.} and not 
all choices of the  
interacting triads give $\Delta E(l)/\Delta E(m)>1$ \cite {Merilees}. 
Nevertheless, Merilees and
Warn \cite {Merilees} find that an initial spectral peak spreads out with most 
of the energy
going upscale---a result very well confirmed numerically. In particular, they 
find that given
an intermediate wavenumber $k$, roughly 70\% of interacting triads exchange more 
energy with
lower wavenumbers, while roughly 60\% of interacting triads exchange more 
enstrophy with higher  
wavenumbers. Given a monoscale forcing at a wavenumber $s$, then, it is tempting 
to arrive at the
picture of the dual cascade of energy to larger scales, and of enstrophy to 
smaller scales.
However, in forced-dissipative turbulence with a full spectrum, the validity of 
this picture is
not at all self-evident.

In particular, the KLB theory envisages an enstrophy dissipation range for 
$k>k_{\nu}$ with
$k_{\nu}\gg\lambda^{1/2}$ (high Reynolds number), so that the forcing and 
dissipation scales are well separated. If molecular viscosity is the only 
dissipation
mechanism, then the dissipation of enstrophy is at least $k_{\nu}^2$ times the 
dissipation of
energy. On the other hand, the forcing of enstrophy is only $\lambda$ times the 
forcing of
energy. This would suggest that forced-dissipative equilibrium is unrealizable 
for
high-Reynolds-number 2D Navier--Stokes turbulence. Yet, Constantin et al. 
\cite{Constantin3,Constantin4} have
proven the existence of a global attractor for this system for the case of 
time-independent forcing.
Unless that case is pathological, it follows that the assumption of an enstrophy 
dissipation range
confined to $k>k_{\nu}\gg\lambda^{1/2}$ must be wrong. In order to deduce the 
directions of energy and
enstrophy cascades, one needs to know the spectral distribution of energy and 
enstrophy dissipation,
and these are not preordained.

\section{Asymptotic behaviour: Dynamical constraint}

We first derive the governing equation for the evolution of $\Lambda$ in the 
absence of a forcing  
term. Taking the scalar product of (\ref{nseqn}) with $u$ and $Au$, 
respectively, we obtain the
governing equations for the decay of the energy and enstrophy for $f=0$ (note 
that the nonlinear
term vanishes in both cases due to the orthogonality properties):
\begin{eqnarray}\frac{\rm d}{{\rm d}t}\norm u^2+2\sum\nu_{\mu}
\norm u_{\mu}^2&=&0,\label{Edecay}\\\frac{\rm d}{{\rm d}t}\norm u_1^2
+2\sum\nu_{\mu}\norm u_{1+\mu}^2&=&0.\label{Zdecay}\end{eqnarray}
Note that the subscript $\mu$ has been dropped from the sum ($\sum$) as there is 
no risk of
confusion. Taking the time derivative of the expression for $\Lambda$ and 
substituting
(\ref{Edecay}) and (\ref{Zdecay}) we obtain
\begin{eqnarray}
\frac{{\rm d}\Lambda}{{\rm d}t}&=&2\sum\nu_{\mu}\frac{\norm u_1^2
\norm u_{\mu}^2-\norm u^2\norm u_{1+\mu}^2}{\norm u^4}\nonumber\\
&=&-2\sum\nu_{\mu}\frac{G(\mu,1,\Lambda,u)}{\norm u^2}.\label{old26}
\end{eqnarray}

A couple of remarks are in order:

\textit{Remark.} If Ekman drag ($\mu=0$) is the only dissipation mechanism 
involved, then ${\rm d}\Lambda/{\rm d}t=0$ because $G(0,1,\Lambda,u)=0$. 
In this special case both enstrophy and energy are dissipated at the same 
rate $2\nu_0$. Since $\Lambda$ remains constant, the distribution of energy 
and enstrophy should then be dramatically different from the viscous
($\mu>0$) cases.

\textit{Remark.} A viscosity causes $\Lambda$ to decrease monotonically since
$G(\mu,1,\Lambda,u)\geq 0$ for all $\mu>0$, while an inverse viscosity has the 
opposite effect. If a single viscosity mechanism is involved, the rate of 
decay of $\Lambda$ is greater for a more spread out spectrum as compared 
with a sharper spectrum having the same energy and enstrophy. This can be 
seen by the fact that $G(\mu,1,\Lambda,u)$ is greater in the former case.

We next consider the time-independent monoscale forcing $f\in 
H(\lambda_s)$, where
$\lambda_s>\lambda_1$, considered by Constantin et al. \cite{Constantin2}.
For the arguments to follow we note that (\ref{nseqn}) possesses a stationary 
solution given by
\begin{eqnarray}
\bar{u}&=&\left(\sum\nu_{\mu}\lambda_s^{\mu}\right)^{-1}f.\label{StatSol}
\end{eqnarray}
This stationary solution is referred to as the primary stationary solution. A
stationary solution other than $\bar{u}$ (if it exists) is identified as a 
secondary stationary solution. The existence of such a solution for the 
traditional 2D Navier--Stokes system under suitable conditions is 
demonstrated in \cite{Iudovich}. The energy and enstrophy evolve according to
\begin{eqnarray}
\frac{1}{2}\frac{\rm d}{{\rm d}t}\norm u^2+\sum\nu_{\mu}\norm
u_{\mu}^2&=&(u,f),\label{Eevolution}\\
\frac{1}{2}\frac{\rm d}{{\rm d}t}\norm
u_1^2+\sum\nu_{\mu}\norm u_{1+\mu}^2&=&(Au,f).\label{Zevolution}
\end{eqnarray}
Since $(Au,f)=\lambda_s(u,f)$ in this case we can multiply the energy equation 
(\ref{Eevolution}) by
$\lambda_s$ and subtract it from the enstrophy equation (\ref{Zevolution}) to 
obtain
\begin{eqnarray}\frac{1}{2}\frac{\rm d}{{\rm d}t}\left\{\norm u_1^2-\lambda_s
\norm u^2\right\}+\sum\nu_{\mu}\left\{\norm u_{1+\mu}^2
-\lambda_s\norm u_{\mu}^2\right\}&=&0.\label{difference}\end{eqnarray}
This equation can be rewritten in terms of the function $G(\mu,1,\lambda_s,u)$ 
as
\begin{eqnarray}\frac{1}{2}\frac{{\rm d}\xi}{{\rm 
d}t}+\sum\nu_{\mu}\lambda_s^{\mu}\xi
&=&-\sum\nu_{\mu}G(\mu,1,\lambda_s,u),\label{xi}\end{eqnarray}
where $\xi\equiv\;\norm u_1^2-\lambda_s\norm u^2$. Now we require that $\mu$ be 
non-negative, i.e. no inverse viscosity is allowed, so that the right-hand side 
of 
(\ref{xi}) is
non-positive. This means that any positive value of $\xi$ will be dissipated 
away, while a
non-positive value will evolve but remain non-positive for all time. To see this 
explicitly,
consider the formal solution of (\ref{xi}), for all $t\geq t_0$, given below:
\begin{eqnarray}
\xi(t)&=&\exp\left\{-2\sum\nu_{\mu}\lambda_s^{\mu}t\right\}
\times\label{xisolution}\\&&\left(\xi_0\exp\left\{2\sum\nu_{\mu}
\lambda_s^{\mu}t_0\right\}-2\int_{t_0}^t\exp\left\{2\sum\nu_{\mu}
\lambda_s^{\mu}t'\right\}\sum\nu_{\mu}G(\mu,1,\lambda_s,u){\rm d}t'\right),
\nonumber
\end{eqnarray}
where $\xi_0=\xi(t=t_0)$. It is easy to see that if $\xi_0\leq 0$ then 
$\xi(t)\leq 0$ for
all time. On the other hand, if $\xi_0>0$ then $\xi(t)$ becomes non-positive for 
all
$t\geq T$ where $T$ is the solution of
\begin{eqnarray}\xi_0\exp\left\{2\sum\nu_{\mu}\lambda_s^{\mu}t_0\right\}&=&
2\int_{t_0}^T\exp\left\{2\sum\nu_{\mu}\lambda_s^{\mu}t'\right\}\sum
\nu_{\mu}G(\mu,1,\lambda_s,u){\rm d}t'.\label{T}\end{eqnarray}
Equation (\ref{T}) always has a finite solution for $T$ unless
$\sum\nu_{\mu}G(\mu,1,\lambda_s,u)$ decreases exponentially in time more rapidly 
than
$\exp\{-2\sum\nu_{\mu}\lambda_s^{\mu}t\}$. Since the function 
$G(\mu,1,\lambda_s,u)$
vanishes if and only if $u\in H(\lambda_s)$, this can occur only if $u$ 
asymptotically approaches
(with respect to the various norms) the eigenspace $H(\lambda_s)$. Now 
$H(\lambda_s)$ is part of
the stable manifold of $\bar{u}=(\sum\nu_{\mu}\lambda_s^{\mu})^{-1}f$; this can 
be seen from
the fact that $B(u,u)=0$ for all $u\in H(\lambda_k),\;\forall k$, so 
$H(\lambda_s)$ is invariant  
and every trajectory on it exponentially converges to $\bar{u}$ at the rate
$2\sum\nu_{\mu}\lambda_s^{\mu}$. It follows that any trajectories which 
asymptotically approach
$H(\lambda_s)$ also converge to $\bar{u}$. Therefore, the cases for which a 
finite solution for $T$
is questionable turn out to be restricted to trajectories on the stable manifold 
of $\bar{u}$.

The argument in the last paragraph indicates that for a general trajectory not 
on the stable manifold
of $\bar{u}$, $\xi$ acquires a non-positive value in a finite time. Within the 
stable manifold of  
$\bar{u}$, only $\bar{u}$ lies on the global attractor and $\xi=0$ for 
$u=\bar{u}$. But before
drawing a final conclusion for $\xi$ on the global attractor it is necessary to 
secure the
non-positiveness of $\xi$ on homoclinic trajectories of $\bar{u}$ which, if they 
exist, are part of
the global attractor. Recall that a homoclinic trajectory (also known as a 
homoclinic orbit) of a
stationary solution is one that emanates from the stationary solution and 
terminates on that same
stationary solution. It is a special limit cycle\footnote{Note that a regular 
limit cycle which would
intersect $H(\lambda_s)$ does not exist because $H(\lambda_s)$ is part of the 
stable manifold of  
$\bar{u}$. Alternatively, the existence of such a cycle would imply from 
(\ref{xisolution}) that
$\int_0^{T_0}\exp\{2\sum\nu_{\mu}\lambda_s^{\mu}t'\}\sum\nu_{\mu}G(\mu,1,\lambda
_s,u)
{\rm d}t'=0$, where $T_0$ is the period of the cycle. But this would require the 
cycle to be
entirely in $H(\lambda_s)$, which would further reduce the cycle to $\bar{u}$.} 
with infinite
period. It can be seen from (\ref{xisolution}) that $\xi$ does not acquire a 
positive value on any
homoclinic orbit of $\bar{u}$ during its infinitely long journey, if such an 
orbit exists. It would,
rather, start from $\bar{u}$ with $\xi=0$, evolve with some negative value for 
$\xi$, and then
$\xi$ would increase to terminate at $\bar{u}$ with $\xi=0$ again.

Therefore, it can be concluded that on the global attractor of the 
Navier--Stokes system, the energy and enstrophy satisfy
\begin{eqnarray}
\norm u^2_1&\leq&\lambda_s\norm u^2.\label{constraint}
\end{eqnarray}
This inequality will be referred to as the dynamical constraint. It is noted, 
for the sake of completeness, that the 
argument leading to
(\ref{constraint}) also implies that $\bar{u}$ is the only point on the global 
attractor where the
equality occurs. Inequality (\ref{constraint}), together with the Poincar\'e 
inequality, gives
\begin{eqnarray}\lambda_1\norm u^2\leq\norm u_1^2\leq
\lambda_s\norm u^2.\end{eqnarray}

An immediate corollary of the dynamical constraint is a similar constraint on 
the enstrophy of a
secondary stationary solution if one exists. In fact, (\ref{constraint}) becomes 
a strict
inequality for any secondary stationary solution. More interpretation of 
(\ref{constraint}) is given
in the next section.

\textit{Remark.} The addition of an inverse viscosity to a system which 
possesses a global
attractor does not jeopardize its existence. However, the spectral distribution 
of energy and
enstrophy on the attractor does not necessarily obey the dynamical constraint in 
this case.

\textit{Remark.} The requirement of a time-independent $f$ is merely for the 
sake of securing the
existence of a global attractor. Equation (\ref{constraint}) is a constraint on 
the long-time
behaviour of the dynamics whether or not the monoscale forcing $f \in H ( 
\lambda_s )$ is 
time-independent.

We now extend the above result to the case of a more general monoscale-like 
forcing as defined
in the Introduction. It is easy to see that when $\lambda$ is constant the above 
analysis applies with
$\lambda$ in place of $\lambda_s$. (In general, there will not exist a 
stationary solution 
as in the case of a monoscale forcing.) For variable $\lambda$, we have the a 
priori inequalities (by hypothesis)
\begin{eqnarray}\lambda_{\min}(u,f)\,\leq\,(Au,f)\,\leq\,\lambda_{\max}(u,f),
~~~(u,f)\geq 0,\label{monoscale-like}\end{eqnarray}
where $\lambda_{\min}$ and $\lambda_{\max}$ are, respectively, the minimum and 
maximum $\lambda_k$ of the
restricted set $K$ of wavenumbers through which energy and enstrophy are 
injected. Note that with variable $\lambda$, we find it necessary to require the 
condition $(u,f)\geq 0$ at all times. Then (\ref{Eevolution}), 
(\ref{Zevolution}), 
and (\ref{monoscale-like}) lead to (\ref{xi}) with the equality replaced
by $\leq$ and with $\lambda_s$ replaced by $\lambda_{\max}$. It follows that 
$\xi (t)$ is bounded from above by zero, for 
sufficiently large time, and thus
the dynamical constraint (\ref{constraint}) must hold on the attractor (if it 
exists) with 
$\lambda_s$ replaced by 
$\lambda_{\max}$.

Finally, if (\ref{monoscale-like}) only holds in a time-mean sense --- note that 
the time-mean energy injection $\langle (u,f) \rangle \geq 0$ necessarily since 
the 
energy dissipation is non-negative --- then we may take the time mean of 
(\ref{xi}) (with $\lambda_s$ replaced by $\lambda_{\max}$ and the equality 
replaced by $\leq$) to obtain $\langle \xi \rangle \leq 0$, equivalently 
\begin{eqnarray}
\langle \norm u^2_1 \rangle \leq \lambda_{\max} \langle \norm u^2 \rangle ,
\end{eqnarray}
and the dynamical constraint holds in the time mean.

\section{Dissipation rates and spectral distributions}

In this section we examine the time-mean dissipation rates of energy and 
enstrophy and 
possible spectral slopes
of energy for three combinations of generalized viscosity: (i) molecular 
viscosity alone,
(ii) molecular
viscosity and Ekman drag, and (iii) molecular viscosity and inverse viscosity. 
Since we are assessing the validity of the KLB theory, which applies to 
statistically stationary turbulence, we may assume that all time-mean quantities 
exist.

For simplicity we first treat the case of monoscale forcing as in the last 
section, leaving the
question
of the applicability of the results to the more general monoscale-like forcing 
until the end of 
the section.
For the first two cases, the results from the last section indicate that both 
energy and its
dissipation are confined to scales no smaller than the injection scale. In 
particular, the dynamical
constraint gives
\begin{eqnarray}
\bar\Lambda \equiv {\langle \norm u^2_1 \rangle \over \langle \norm u^2 \rangle} 
\leq\lambda_s.\label{constraint1}
\end{eqnarray}
This means that in both cases the spectrum adjusts so that the dissipation of 
energy takes place on
scales no smaller than the forcing scale. Moreover, the energy dissipation rates 
are bounded from
above by $2\nu_1\lambda_s$ for case (i) and by $2\nu_1\lambda_s+2\nu_0$ for case 
(ii). 

We know that viscosity acts on smaller scales for enstrophy than it does for 
energy. The question is
where, in wavenumber space, the enstrophy dissipation takes place. To answer 
this question we
resort to (\ref{difference}). Since both energy and enstrophy are bounded from 
above on the global
attractor, taking the time mean of (\ref{difference}) leads to
\begin{eqnarray}
\sum\nu_{\mu}\left\{
\langle \norm u_{1+\mu}^2 \rangle -\lambda_s \langle \norm u_{\mu}^2 \rangle 
\right\}&=&0.\label{balance}
\end{eqnarray} 
This equation is referred to as the balance equation. Attempts to constrain the 
time-mean enstrophy dissipation rate and the scaling law of the energy spectrum 
constitute the 
remainder of this section.

~

{\bf Case (i):} The balance equation reads in this case
\begin{eqnarray}
\langle \norm u_2^2 \rangle = \lambda_s \langle \norm u_1^2 \rangle 
,\end{eqnarray}
so
\begin{eqnarray}\bar\Lambda_{Z1} \equiv {\langle \norm u_2^2 \rangle \over 
\langle \norm u_1^2 \rangle} = \lambda_s .\label{stat1}\end{eqnarray}
This well-known result corresponds to (6) of Constantin et 
al.~\cite{Constantin2}. It implies that the time-mean
enstrophy
dissipation rate is given by $2\nu_1\lambda_s$, and that the enstrophy 
dissipation is concentrated
around the forcing scale. It follows that there can be no direct enstrophy 
cascade. Moreover, this
constraint gives rise to a very steep spectral slope for $k>s$. Let $E(k)$ be 
defined by
\begin{eqnarray}E(k)&=&\langle 
\norm{P(\lambda_k)u}^2\rangle\nonumber\end{eqnarray}
so that the total energy $\mathcal{E}$ is given by
\begin{eqnarray}\mathcal{E}&=&\int_0^{\infty}E(k){\rm 
d}k,\nonumber\end{eqnarray}
where the discrete wavenumber has been changed to a continuous one for 
the sake of convenience.
Now suppose that $E(k)\propto k^{-\delta}$ over a range $[k_0,k_1]$ where 
$s<k_0\ll k_1$; the
dissipation of enstrophy over that range, $D_Z$, is then given by
\begin{eqnarray}D_Z&\propto&\int_{k_0}^{k_1}k^{4-\delta}{\rm 
d}k\,=\,\frac{1}{5-\delta}
\left(k_1^{5-\delta}-k_0^{5-\delta}\right)~~~[\delta\neq 5].\end{eqnarray}
Equation (\ref{stat1}) requires that $D_Z$ be dominated by the large scales. As 
a result, the
satisfactory values of $\delta$ are confined to $\delta>5$. ($\delta=5$ is not 
acceptable.) 

\textit{Remark.} A scaling law of the energy spectrum $k^{-\delta}$ with 
$\delta>5$ is much steeper
than the classical $k^{-3}$ inertial range power law (see Kraichnan 
\cite{Kraichnan1,Kraichnan2}). Hence, the
result of this section suggests that the latter scaling law is unrealizable in 
2D Navier--Stokes
turbulence for a constant monoscale forcing. This result was derived by 
Constantin et al.~\cite{Constantin2}.   

\textit{Remark.} For a viscosity of degree $\mu\geq 0$ instead of molecular 
viscosity, which   
includes Ekman drag ($\mu=0$), the satisfactory values of $\delta$ are confined 
to $\delta>3+2\mu$.

\textit{Remark.} Although an inverse energy cascade cannot be excluded, we can 
rule out the classical
$k^{-5/3}$ power law \cite{Kraichnan1} for any such range by a similar argument. 
Suppose that
$E(k)\propto k^{-\gamma}$ over a range $[k_0,k_1]$ where $k_0\ll k_1<s$; the
dissipation of energy over that range, $D_E$, is then given by
\begin{eqnarray}D_E&\propto&\int_{k_0}^{k_1}k^{2-\gamma}{\rm 
d}k\,=\,\frac{1}{3-\gamma}
\left(k_1^{3-\gamma}-k_0^{3-\gamma}\right)~~~[\gamma\neq 3].\end{eqnarray}
An inverse cascade of energy requires that $D_E$ be dominated by the small-$k$ 
end of the range,
which
requires $\gamma>3$. ($\gamma=3$ is not acceptable.) Therefore, the $k^{-5/3}$ 
power law for 2D
Navier--Stokes turbulence is incompatible with an inverse energy cascading 
range. This result is clearly independent of the nature of the forcing.

~

{\bf Case (ii):} The balance equation reads in this case
\begin{eqnarray}
\nu_1\left\{\langle\norm
u_2^2\rangle-\lambda_s\langle\norm u_1^2\rangle\right\}+\nu_0\left\{\langle\norm
u_1^2\rangle-\lambda_s\langle\norm u^2\rangle\right\}&=&0,
\end{eqnarray}
so
\begin{eqnarray}
\nu_1\bar\Lambda_{Z1}+\nu_0&=&\nu_1\lambda_s+\frac{\nu_0\lambda_s}
{\bar\Lambda}\,\geq\,\nu_1\lambda_s+\nu_0.\label{Case2}
\end{eqnarray}
This implies that the time-mean enstrophy dissipation rate is bounded from below 
by
$2\nu_1\lambda_s+2\nu_0$. Equivalently, we have $\bar\Lambda_{Z1}\geq\lambda_s$. 
In 
other words,
Ekman drag allows the spectrum to adjust toward the small scales (as compared 
with case (i) for
which $\bar\Lambda_{Z1}=\lambda_s$), so that the enstrophy dissipation can occur 
at 
scales smaller than  
the forcing scale. To see the extent of such an adjustment we rewrite the 
equality in (\ref{Case2}) as
\begin{eqnarray}\frac{\bar\Lambda_{Z1}}{\lambda_s}&=&1+\frac{\nu_0}{\nu_1\lambda
_s}
\left(\frac{\lambda_s}{\bar\Lambda}-1\right).\label{lambdaratio}\end{eqnarray}
The ratio $\bar\Lambda_{Z1}/\lambda_s>1$ because $\lambda_s/\bar\Lambda>1$ 
(strictly 
speaking we need to
exclude the primary stationary solution $\bar{u}$, the only point on the global 
attractor for which
$\lambda_s/\bar\Lambda=1$). The question is whether $\bar\Lambda_{Z1}/\lambda_s$ 
can be 
much greater than
unity
for a system with $\nu_0\gg\nu_1\lambda_s$, where the large-scale dissipation is 
dominated by Ekman
drag. This is not obvious since we expect that $\lambda_s/\bar\Lambda\rightarrow 
1$ 
as
$\nu_0/(\nu_1\lambda_s)\rightarrow\infty$ because the dynamical constraint is 
$\bar\Lambda=\lambda_s$ when
Ekman drag alone is responsible for dissipation. However, we do not know exactly 
how this limit is
approached as $\nu_0/(\nu_1\lambda_s)\rightarrow\infty$. Nevertheless, it 
appears that in principle a
direct enstrophy cascade could be achieved in this system.   

We now examine implications for scaling of the energy spectrum. For an energy 
cascading range $k_0\ll
k_1<s$, the energy dissipation is given, with the same notation as before, by
\begin{eqnarray}
D_E&\propto&\int_{k_0}^{k_1}\left(\nu_0k^{-\gamma}+\nu_1k^{2-\gamma}\right)
{\rm 
d}k\nonumber\\&=&\frac{\nu_0}{1-\gamma}\left(k_1^{1-\gamma}-k_0^{1-\gamma}\right
)
+\frac{\nu_1}{3-\gamma}\left(k_1^{3-\gamma}-k_0^{3-\gamma}\right)~~~
[\gamma\neq 1,3].
\end{eqnarray}
We have already assumed that $\nu_0\gg\nu_1\lambda_s$ so that Ekman drag 
dominates on the large scales,
and therefore the energy dissipation occurs primarily on the large scales 
provided $\gamma>1$. On the
other hand, we want negligible dissipation of enstrophy on the large scales, and 
this requires
$\gamma<3$. Thus $1<\gamma<3$, which includes the $\gamma=5/3$ power law of the 
KLB theory.

For an enstrophy-cascading range $s<k_0\ll k_1$, the enstrophy dissipation is 
given by  
\begin{eqnarray}
D_Z&\propto&\int_{k_0}^{k_1}\left(\nu_0k^{2-\delta}+\nu_1k^{4-\delta}\right)
{\rm 
d}k\nonumber\\&=&\frac{\nu_0}{3-\delta}\left(k_1^{3-\delta}-k_0^{3-\delta}\right
)
+\frac{\nu_1}{5-\delta}\left(k_1^{5-\delta}-k_0^{5-\delta}\right)~~~[\delta\neq 
3,5].\end{eqnarray}
It is seen that enstrophy dissipation will occur primarily on the small scales 
provided
$\delta<5$. ($\delta=5$ is not acceptable.) On the other hand, we want 
negligible dissipation of
energy
on the small scales, and this requires $\delta>3$. ($\delta=3$ is not 
acceptable.) Thus $3<\delta<5$,  
and the $\delta=3$ power law of the KLB theory (with or without the logarithmic 
correction of
\cite{Kraichnan2}) is not permitted. In fact, it is in any case excluded (as in 
case (i)) by the
dynamical constraint~(\ref{constraint1}).

~

{\bf Case (iii):} Similar to case (ii) we have
\begin{eqnarray}\frac{\bar\Lambda_{Z1}}{\lambda_s}&=&
1+\frac{\nu_{\mu}\bar\Lambda_{E\mu}^{\mu}}
{\nu_1\lambda_s}\left(\frac{\lambda_s}{\bar\Lambda}
-\frac{\bar\Lambda_{Z\mu}^{\mu}}
{\bar\Lambda_{E\mu}^{\mu}}\right),\label{stat4}\end{eqnarray}
where
\begin{eqnarray}
\bar\Lambda_{E\mu}^{\mu} \equiv \frac{\langle\norm 
u_{\mu}^2\rangle}{\langle\norm u^2\rangle} ,\quad \bar\Lambda_{Z\mu}^{\mu} 
\equiv \frac{\langle\norm u_{1+\mu}^2\rangle}{\langle\norm 
u_1^2\rangle} .\end{eqnarray}
Unlike case (ii) where only one ratio on the right-hand side of 
(\ref{lambdaratio}), namely
$\lambda_s/\bar\Lambda$, depends on the actual spectrum, all three ratios on the 
right-hand side
of (\ref{stat4}) are spectrum dependent. Among them we expect
$\bar\Lambda_{Z\mu}^{\mu}/\bar\Lambda_{E\mu}^{\mu}$ to be smaller than unity for 
negative
$\mu$ (see section 2). Now the ratio $\lambda_s/\bar\Lambda$ is not always 
greater 
than unity, and
depends on the outcome of the competition between molecular viscosity and 
inverse viscosity.
Nevertheless, for suitable values of $\nu_{\mu}$ and $\mu$ a dual cascade 
certainly appears to be
possible. In this case, energy cascades upscale and is dissipated by inverse 
viscosity, while
enstrophy
cascades downscale and is dissipated by molecular viscosity. Of course, inverse 
viscosity has no
physical basis and is employed purely for numerical reasons.

We now extend the results of this section to the more general case of a 
monoscale-like forcing. It is only necessary to do so for cases (i) and (ii) 
since those are where we
have derived restrictions. It is easy to see that everything goes through when 
$\lambda$ is
constant. For variable $\lambda$, or if the condition of monoscale-like forcing 
holds only in the time mean, the balance equation (\ref{balance}) may be 
replaced by
\begin{eqnarray}\sum\nu_{\mu}\left\{
\langle\norm
u_{1+\mu}^2\rangle-\lambda_{\max}\langle\norm 
u_{\mu}^2\rangle\right\}&\leq&0.\label{balance1}\end{eqnarray}
The arguments concerning admissible spectral slopes go through without change 
since they do not refer
at
all to the forcing mechanism; in the definitions of the inertial subranges the 
forcing wavenumber
$s$ is replaced with either $\lambda_{\min}^{1/2}$ or $\lambda_{\max}^{1/2}$, as 
appropriate. This
leaves
only the prohibition of a direct enstrophy cascade in case (i). There one 
obtains $\bar\Lambda_{Z1} \leq
\lambda_{\max}$ in place of (\ref{stat1}), which still precludes a direct 
enstrophy cascade.

\section{Concluding remarks}

In this paper we have derived various constraints on the asymptotic 
behaviour of the 2D Navier--Stokes equations in a finite 
domain driven by a 
monoscale-like 
forcing and damped by a general class of dissipation operators. By 
monoscale-like forcing we mean that the ratio of enstrophy to energy 
injection lies within the range of the square of the forcing 
wavenumbers, for all possible velocity fields $u$, which is a common (though not 
exclusive) scenario in the 
KLB theory (e.g. Kraichnan \cite{Kraichnan1}, p.1421b; Pouquet et al. 
\cite{Pouquet}, 
p.314; and Lesieur \cite{Lesieur}, p.291). Note that this crucial property is 
{\it not} 
guaranteed for a constant external forcing, except in the special case of a 
monoscale forcing. The results obtained include constraints on the time-mean 
enstrophy dissipation 
range, an upper
bound on the system's enstrophy in terms of the energy and the forcing scale, 
and bounds on possible
scaling laws for the energy spectrum. If the condition of monoscale-like forcing 
holds only in the time mean, then the upper bound on the enstrophy holds in the 
time mean rather than asymptotically. The validity of the dual cascade picture 
is examined by
determining where, in wavenumber space, dissipation of energy and enstrophy 
occurs.

For 2D Navier--Stokes turbulence, with molecular viscosity the only form of 
dissipation, no direct   
enstrophy cascade is permitted; rather, enstrophy dissipation is required to 
occur in the vicinity
of the forcing scale. Although a reverse energy cascade is permitted (not to say 
that it occurs), it is 
shown to be incompatible
with the $-5/3$ power-law scaling of KLB theory for the energy spectrum in such 
a range (as noted previously by Constantin et al.~\cite{Constantin2}).

Ekman drag has often been employed in numerical simulations of 2D turbulence to 
allow a second
dissipation channel. Furthermore, Ekman drag is a reasonable representation of a 
frictional
planetary boundary layer in the geophysical context. We show rigorously how the 
introduction of
Ekman drag together with viscosity breaks the strong constraints on the spectral 
distribution of
enstrophy dissipation, and thereby allows the possibility of a dual cascade. 
However, the rigorous
bound on the enstrophy precludes the existence of the $-3$ power-law scaling of 
KLB theory (with or
without a logarithmic correction) for the energy spectrum in an 
enstrophy-cascading range. Instead,
the power law is shown to be between $-3$ and $-5$. This is consistent with the 
published results of
numerical simulations.

It is shown that the use of an inverse viscosity together with viscosity allows 
the possibility of
KLB power-law scaling in the enstrophy-cascading range. Indeed, the numerical 
simulations of Borue
\cite{Borue}, and the more recent high-resolution numerical simulations of 
Lindborg and Alvelius
\cite{Lindborg}, which both claim to exhibit this scaling, employ an inverse 
viscosity. Our results
suggest that this is not a coincidence.

It should be emphasized that our derived constraints on power laws for the 
energy 
spectrum apply for any spectrally localized forcing, not just for  
monoscale-like forcing.

The results derived in this paper are believed to be particular to 
2D turbulence, because they all rely on the conservation of both 
energy and enstrophy in nonlinear interactions, as expressed through 
(\ref{Eevolution}) and (\ref{Zevolution}) and subsequent key relations 
such as (\ref{difference}), 
(\ref{balance}), and (\ref{balance1}). There is no analogue of 
(\ref{Zevolution}) in 3D turbulence.

Constantin et al.~\cite{Constantin2} have shown that for the special case 
of a spectrally-localized constant external forcing, the realization of 
a direct enstrophy cascade in forced 2D Navier--Stokes turbulence in a finite 
domain requires a combination of 
energy input and energy removal by the forcing. The generalization of this 
result to our case is that a direct enstrophy cascade within forced 2D 
Navier--Stokes 
turbulence requires the abandonment of the popular concept of monoscale-like  
forcing (as defined here). Of course, the $-5/3$ power-law  scaling remains 
incompatible with an upscale energy  cascade, irrespective of the forcing 
mechanism.

A general result of this study is that the choice of forcing mechanisms and 
dissipation operators has implications for the
spectral distribution of energy and enstrophy dissipation, and thus for the 
possible existence of energy and enstrophy cascades. Furthermore, the choice 
of dissipation operators has implications for permissible scalings of the 
energy spectrum. In choosing forcing mechanisms and dissipation operators 
for numerical reasons, one should be mindful of these constraints.

{\bf Acknowledgements}

The work reported here represents part of CVT's Ph.D. thesis at the 
University of Toronto, which was supported by University of Toronto 
Open Fellowships and Department of Physics Burton Fellowships. TGS would 
like to acknowledge support from the Natural Sciences and Engineering
Research Council and the Meteorological Service of Canada. Both authors are 
grateful to John Bowman, Joe Tribbia, Jeff Weiss, and particularly Ricardo Rosa 
for helpful comments on the manuscript.

\end{document}